\documentclass[onecolumn,preprintnumbers,amsmath,amssymb]{revtex4}
\usepackage{graphicx}% Include figure files
\begin{document}
%\preprint{APS/123-QED}

\title{Bose-Einstein condensation on an atom chip}
%\footnote{Supported by the State Key Basic Research Program (Grant No.2001CB309307)}

\author{Yan Bo$^1$, Cheng Feng$^1$, Ke Min$^1$, Li Xiaolin $^1$, Tang Jiuyao$^2$}
 %\altaffiliation{}%Lines break automatically or can be forced with \\
\author{Wang Yuzhu$^1$}%
 \email{yzwang@mail.shcnc.ac.cn}
\affiliation{$^1$ Key Laboratory for Quantum Optics, Center for Cold
Atom Physics, Shanghai Institute of Optics and Fine Mechanics,
Chinese Academy of Sciences, Shanghai, 201800\\ $^2$ Department of
Physics, Zhejiang University, Hangzhou, 310027
}%

\date{\today}

\begin{abstract}
We report an experiment of creating Bose-Einstein condensate (BEC)
on an atom chip. The chip based Z-wire current and a homogeneous
bias magnetic field create a tight magnetic trap, which allows for a
fast production of BEC. After an 4.17s forced radio frequency
evaporative cooling, a condensate with about 3000 atoms appears. And
the transition temperature is about 300nK. This compact system is
quite robust, allowing for versatile extensions and further studying
of BEC.
PACC: {3280P, 4250}
\end{abstract}
%\pagestyle{myheadings} \markright{sadf}
%\pacs{42.50.Vk,03.75.Be,32.80.Pj}
\maketitle
\section*{1. Introduction}
The realization of Bose-Einstein condensation is a great achievement
in physics \cite{science BEC,Science BEC2}. Although many groups in
the world have produced BEC, making BEC is still a challenge to a
new group \cite{siomBEC}. Now, BEC has been produced in several type
of systems, such as magnetic trap \cite{quic trap}, dipole trap
\cite{dipole trap}, mini trap \cite{minitrap1,minitrap2} and atom
chip \cite{zimmermann,hansch}. Among them, atom chip is specially
intriguing. It allows for a fast production of BEC since it provides
a much tighter trap compared with the traditional magnetic trap. A
tighter trap means a higher elastic collision rate, which leads to a
less strict vacuum requirement. A pressure about $10^{-8}$ Pa is
enough. Also, the wires on the atom chip can be designed flexibly to
meet different goals, like Ioffe-Pritchard trap \cite{chipreview},
magnetic lattice \cite{lattice} and double traps \cite{split}. In
addition, optical components can be easily integrated on an atom
chip, such as micro-cavity \cite{cavity}, optical fibre \cite{fiber}
and so on. Atom chip is an useful platform to study BEC extensively.
In this paper, we report our experiment of making BEC on an atom
chip.

\section*{2. Our experimental setup}
Our experimental system is a single-chamber system, as shown in Fig.
1. A quartz cell of $40mm\times40mm\times120mm$ is used as the main
working chamber, a chip is placed up side down in the cell. The atom
source is provided by a $^{87}Rb$ dispenser which can be controlled
by electric current conveniently. A background pressure below
$1\times10^{-8}$ Pa is maintained  by an ion pump and a sublimation
pump.

For the optical system, we use three diode lasers to form our four
laser beams: the trapping light, the repumping light, the pumping
light and the probe light. All lasers are locked to the saturated
absorption spectrums, then are modulated by acousto-optical
modulations (AOM) to the right frequencies and intensities. Finally
they are coupled into four single mode fibers and injected to the
cell. The trapping light and the repumping light are used in
magneto-optical trap (MOT), the pumping light is used to pump atoms
to $m_F=2$ state which increases the efficiency of loading atoms to
the chip-based Z-trap.

Our chip is shown in Fig. 2. A gold layer with a thickness of $5\mu
m$ is deposited on a $Si$ wafer with a 500nm $SiO_2$ film. The wires
are defined by photolithography which produces gaps to insulate the
wires from the rest gold layer. The width of the gap is about $10\mu
m$. The surface of the gold layer is excellently smooth with a
$90\%$ reflection at 780nm wavelength when the incident angle is
$45^\circ$. Since we use the gold layer as a mirror to reflect the
laser beams to form a mirror-MOT \cite{hansch}. Good reflection
increases atom number collected by the mirror-MOT. There are two
U-shaped wires with width of $200\mu m$, a Z-shaped wire with width
of $100\mu m$ and a W-shaped wire on the chip. More details of the
chip fabrication can refer to  \cite{lixiaolin,yanbo}. The Z-wire
combining with a bias field create an Ioffe-Pritchard-like trap
which has no zero-field point. It is useful for evaporative cooling
since the "Majorana flops" can be avoided \cite{metcalf}. And we use
the W-wire as the radio frequency (RF) terminal for RF-evaporative
cooling, because it is near to the Z-wire, which avoids the shield
effect of the gold layer. Many researchers show that gold layer
absorbs Rb atoms, which leads to a reduction of the light
reflectivity and pollutes the vacuum \cite{Dushengwang}. In order to
overcome this disadvantage, a thin $10nm$ quartz film is deposited
on the chip surface. This thin film does not effect the reflectivity
since the thickness is much smaller than the laser wavelength, and
can isolate the strong attractive interaction between Au and Rb
atoms.

\section*{3. Vacuum and lifetime}
The preconditions of creating BEC is extremely strict. One of the
most important demands is ultra-high vacuum (UHV). Even for an atom
chip system, where the vacuum limitation is a bit loose, a pressure
of $10^{-8}$ Pa is still required. So special attentions should be
paid to the vacuum system. All the components in the vacuum are UHV
compatible and pretreated in strictly clean processes. After
installed, the whole system is baked at $120^\circ C$ for 72 hours
continuously.  And the vacuum of our system can reach about
$1\times10^{-8}$ Pa. It is maintained by a $40 L/s$ ion pump and a
sublimation pump which works once a week. We also place a getter in
the cell to improve the pump rate. It is a passive pump by absorbing
gases and works even in the room temperature.

Our atom source is a Rb dispenser. When heated by an electric
current, it releases Rb atoms. Such a device provides a convenient
way for time-dependent control of Rb pressure. It is useful for
achieving a UHV environment \cite{dispenser,dispenser2}. The
critical current for our dispenser starting to release atoms is
about $3.5A$. In the experiment of BEC, we use a pulsed method:
heating the dispenser with a $5.7A$ current pulse for $30s$, then
waiting for $60s$ to recover the vacuum, and then starting the MOT
to collect cold atoms. In this way, adequate atoms are collected in
MOT while a relative good vacuum pressure is obtained. In our
experiment, typically $2\times10^7$ atoms are collected in MOT, and
$3.5\times10^6$ atoms are loaded to the Z-trap, the temperature of
atoms in the Z-trap is about $100\mu K$, the lifetime is $11s$. It
is the beginning of the forced evaporative cooling.
\section*{4. Forced radio frequency evaporative cooling and Bose-Einstein condensate}
Up to now, the evaporative cooling is the only way to achieve BEC
\cite{makingBEC}. In order to evaporate effectively, a runaway
evaporative cooling regime should be reached. The elastic collision
rate $\gamma_{el}$ and the lifetime of atoms in the Z-trap
$\tau_{loss}$ should satisfy \cite{goodtobad}:
\begin{equation}
\gamma_{el} \tau_{loss} >150,
\end{equation}
which is usually called good-to-bad collision ratio. In order to
increase the elastic collision rate, we adiabatically compress the
Z-trap to $I=3.6A, B_x=4G, B_y=72G$, corresponding to a trap
frequency of $(26, 3900, 3900) Hz$. It is about $90\mu m$ away from
the chip surface. The measured good-to-bad collision ratio is
$\gamma_{el} \tau_{loss}
>1000$. Runaway evaporation regime is reached. After an optimized
3.5s RF-evaporative cooling from $74M$ to $4.3M$, $1.3\times10^5$
atoms are left, the temperature is about $10\mu K$. In order to
reduce the heating rate and three-body collisions, we decompress the
trap to $I=2A, B_x=2.8G, B_y=33.6G$, corresponding to $(21, 1900,
1900) Hz$. It is about $110\mu m$ away from the chip surface. This
decompress process lasts $160ms$, and at the same time a RF sweep
from $4.3M$ to $2.7M$ is present. The atom cloud is further cooled,
$1.1\times10^5$ atoms are left and the temperature drops to $3.4\mu
K$. Finally, we apply an RF sweep from $2.7M$ to $2.315M$, lasting
$510ms$, and BEC appears. We use the time of flight method to
determine the result of phase transition. In order to get a good
signal-to-noise ratio, we decompress the Z-trap before releasing the
atoms. Atoms are decompressed to a trap of $I=3.6A, B_x=0, B_y=24G$,
the trap frequency is $(38, 700,700)Hz$. The adiabatic decompression
does not change the phase space density but reduces the temperature
and moves atoms away from the chip surface, so a better signal to
noise ratio is obtained. As shown in Fig. 3. The nonisotropic
expansion of the atom cloud in free space is observed which is a
strong evidence of achieving BEC. The flight time is $2ms, 6ms,
10ms, 14ms$ and $20ms$ respectively. The transition temperature is
about $300nK$, and there are about 3000 atoms in the pure BEC.

The aspect ratio after atoms are released from the Z-trap is
measured. For a thermal cloud, it expands isotropically, the aspect
ratio approaches one as the flight time becomes longer. For a BEC
released from a cigar shape trap, the radius along the tightly
trapped direction expands rapidly. This is a fundamental difference
between BEC and thermal gases \cite{expansion1,expansion2}. We
observe this phenomena when we detect the time of flight signal as
shown in Fig. 4. The aspect ratio becomes much bigger than one when
the flight time is longer than 10ms. The change of aspect ratio
during free expansion is also an important criteria of achieving
BEC.

In order to measure the critical transition point of our BEC, we
stop the RF sweep at different frequency. As shown in Fig. 5, the
pictures are taken after 12ms of flight. When the end frequency is
$2.36M$ or $2.34M$, they are still thermal gases since the profiles
of optical depth are Gauss distributions. But when we sweep to
$2.33M$, a bimodal distribution appears, the optical depth increases
suddenly and the cloud size reduces suddenly. At this time, a
partial condensate appears. When we sweep further to $2.32M$, It is
clearly shown that a sharp peek lies on a low thermal gas
background. At this time, the BEC is quite pure. There are some
stripes shown in the photo when BEC appears, It is due to the
diffraction of the imaging system \cite{quqiuzhi}.

We also measure the lifetime of BEC in the Z-trap. The dominating
limitation of the lifetime is the heating rate when atoms are held
in the trap. We measure the temperature when atoms are held in the
Z-trap for different time. A heating rate of $3.4\mu K/s$ is
detected. The thermal radiation of the chip, the stray light
scattering and the current noise are all possible heating sources
\cite{chipreview}. In this condition, the measured lifetime of BEC
is about 90ms. In order to increase the lifetime, a RF field at the
end frequency is present when atoms are held. In this way, heated
atoms are evaporated away by RF radiation before interacting with
other atoms and do not destroy the BEC. And the lifetime is
prolonged to 500ms.

\section*{6. Conclusion and acknowledge}
In conclusion, We have achieved BEC on our atom chip system. As we
know, it is the first chip BEC in China. The nonisotropic expansion,
bimodal distribution profile and aspect ratio measurement after
releasing provide the strong evidences of BEC creation.  Such a fast
production of BEC (evaporative cooling time is less than 5 seconds
and the circle time is about 2 minutes) provides a robust and
convenient platform to study the properties of this special matter.
Further experiments are being carried out and we hope to go deeper
into the research of BEC.

We acknowledge Wei rong, Zhou Shuyu and Xu Zhen for helpful
discussions, Zhou Shanyu, Xu Daming and Wu Haibing for technical
supports. The project is supported by the National Basic Research
program of China under Grant No. 2006CB921202.
\begin{figure}
\includegraphics[height=6cm]{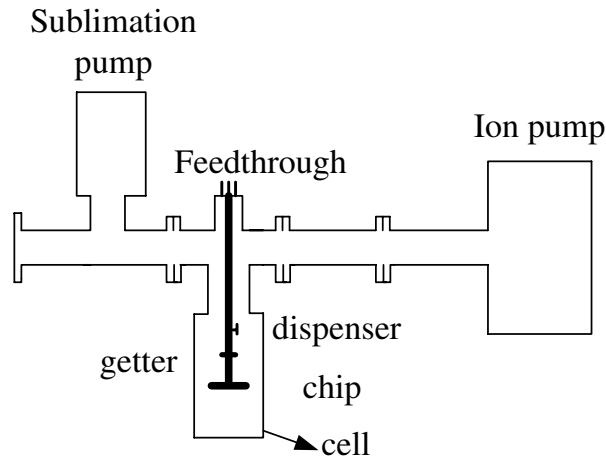}
\caption{The schematic setup of our vacuum system. It is a
single-champer system with a quartz cell. An ion pump, a passive
getter and a sublimation pump are used to keep the vacuum. The atom
source is a dispenser.}
\end{figure}

\begin{figure}
\includegraphics[height=6cm]{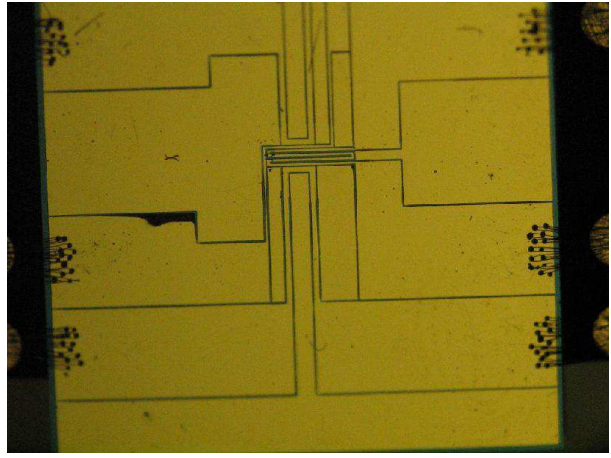}
\caption{The picture of our chip. The gold layer is $5\mu m$
thickness. The wires are isolated from the rest gold layer by $10\mu
m$ gaps. The width of two U-wires is $200\mu m$, and $100\mu m$ for
the Z-wire. Each chip wire is connected to the electrode by spot
welding of 20 thin gold wires as shown in the picture.}
\end{figure}

\begin{figure}
\includegraphics[height=6cm]{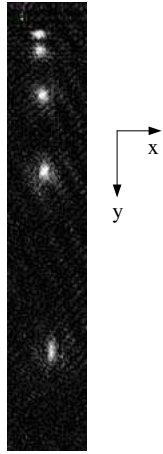}
\caption{Time of flight images of the atom cloud after released from
the trap. The free flight time is 2ms, 6ms, 10ms, 14ms and 20ms
respectively from top to bottom. The nonisotropic expansion is a
strong evidence of BEC.}
\end{figure}

\begin{figure}
\includegraphics[height=6cm]{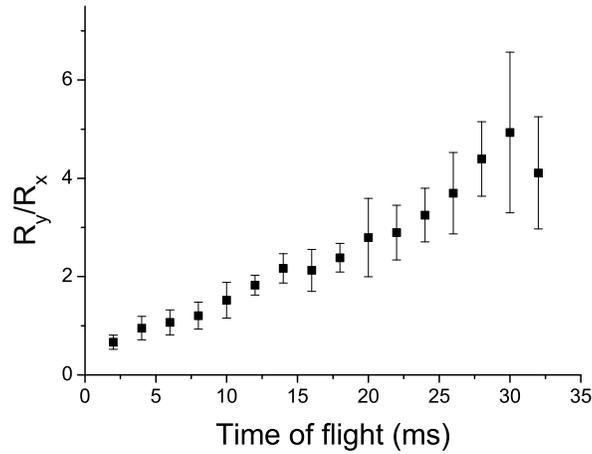}
\caption{The measurement of aspect ratio of free expansion between
the cloud radius in y direction and x direction as shown in Fig. 3.
The radius in the tight trapped direction expands rapidly, and the
aspect ratio is much bigger than one when time of flight is longer
than 10ms. It is an important criteria of achieving BEC.}
\end{figure}

\begin{figure}
\begin{minipage}[c]{\textwidth}
\centering
\includegraphics[height=6cm]{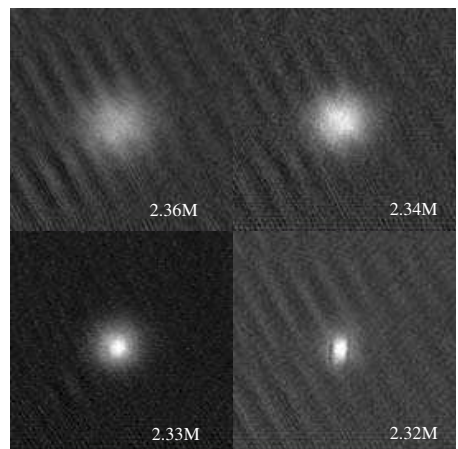}
\end{minipage}
(a)
\begin{minipage}[c]{\textwidth}
\centering
\includegraphics[height=6cm]{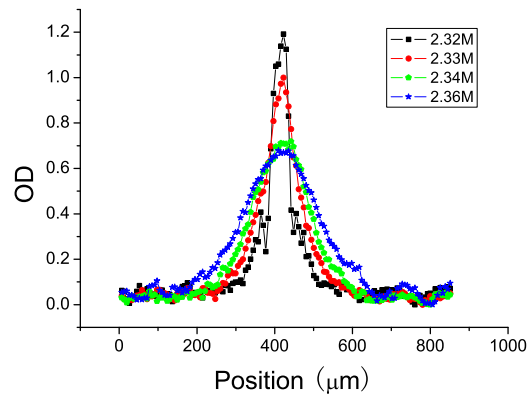}
\end{minipage}
(b) \caption{Determining the critical frequency where BEC appears.
(a) When the end frequency is 2.34MHz, it is still a thermal cloud,
and the profile is a Gauss distribution. When end at 2.33MHz,
partial condensate appears. And when end at 2.32M, a more pure BEC
is created and a sharper peak of the profile appears.(b) is the
section profile of the cloud. A sudden increasing of optical depth
and a bimodal distribution appear when sweep to 2.33MHz which is the
critical frequency when BEC appears.}
\end{figure}

\end{document}